\documentclass[12pt]{iopart}
\usepackage[dvips]{graphicx}
\begin{document} 
\letter{Spin dependent quantum transport effects in Cu nanowires} 

\author{DM Gillingham, C M\"{u}ller, JAC Bland}
\address{Cavendish Laboratory, University of Cambridge, Cambridge, CB3~0HE, UK}

\eads{jacb1@phy.cam.ac.uk}

\begin{abstract}
In this work we investigate quantum transport in Cu nanowires created by bringing macroscopic Cu wires into and out of contact under an applied magnetic field in air.  Here we show that a 70 \% magneto-conductance effect can be seen in a Cu nanowire in a field of 2 mT at room temperature. We propose that this phenomenon is a consequence of spin filtering due to the adsorption of atmospheric oxygen modifying the electronic band structure and introducing spin split conduction channels.  This is a remarkable result since bulk Cu is not magnetic and it may provide a new perspective in  the quest for spintronic devices. 
\end{abstract}

\submitto{\JPCM}
\maketitle

\maketitle

The very rapidly developing field of spintronics\cite{prinz1999} is based on the proposal to use the spin of the electron instead of its electrical charge in future devices.  The challenge is to achieve efficient spin injection at room temperature and to conclusively demonstrate the manipulation of electron spin in an all--electrical manner.  Spintronics requires small structures to satisfy the requirement that the spins act coherently and that the electrons travel ballistically \cite{schmidt2000,rashba2000}.  Furthermore, low dimensional structures such as nanowires can have vastly different electronic structure compared to that of the bulk material and this can, on a sufficiently small scale be expected to be highly significant in the operation of spintronic devices.   Here we present the observation of magnetic field dependent quantum transport in Cu nanowires. A large magneto-conductance effect arises, which we attribute to the formation of spin split conduction states in the nanowire. 

A nanowire can be considered to be an atomic sized constriction between two electron reservoirs.  If such a nanowire is small enough, the electrical conduction $(G)$ can be quantized according to the Landauer equation\cite{Yacoby1990,glazman1988,brandbyge1997} 
\begin{equation}
G = G_0\sum_{n\sigma}T_{n\sigma}
\label{eqn:landuer}
\end{equation}
where $G_0=e^2/h$ is the conductance quantum ($e$ is the electronic charge, $h$ is Planck's constant), $T_{n\sigma}$ is a transmission coefficient for the n$^{th}$ channel and electron spin $\sigma$ (which can take one of two values either $\uparrow$ or $\downarrow$).  For ballistic transport $T_{n\sigma}$ can either be 1 or 0 corresponding to an open or closed channel. For non-magnetic materials one would expect the different spin channels to have the same energy and to be degenerate and in this case $G_0=2e^2/h$.  It has been shown that for a two dimensional electron gas \cite{thomas2000} and ferromagnetic metals (such as Fe and Ni) \cite{Garcia2000,komori2001,ono1999,oshima1998} that this spin degeneracy can be lifted.  Garcia et al, have reported very large magneto-conductance effects of the order of 300 \% in Ni nanowires\cite{tatara1999}; however Ni is a ferromagnetic metal and so magneto-conductance effects can be expected in this case.  
\begin{figure}\begin{center}
    \includegraphics[width=10cm]{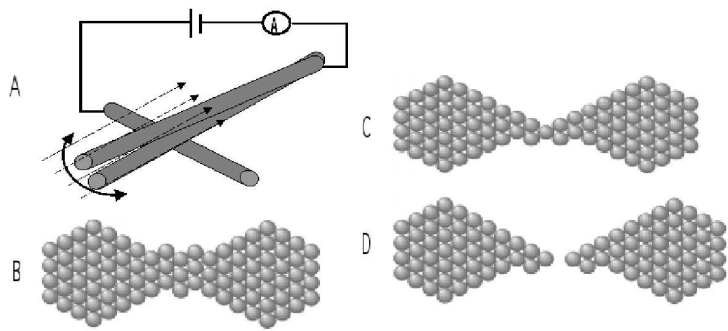}
    \caption{An outline of the experimental system used.  (A) is a schematic of the equipment, the nanowire is made where the wires come into contact. The magnetic field is applied in the direction of the dashed arrows. (B)-(D) is an illustration of how the nanowires are created.  When the macroscopic wires are in contact atoms bind to both wires (B), as the wires separate the metal forms a neck which stretches out till it is atomic sized (C) - then it snaps (D).  The quantum conduction is seen in region (C) close to (D) where the wires snap. }
\end{center}\end{figure}

The nanowires were made by tapping two macroscopic Cu wires together in air at room temperature\cite{Costa-Kramer1995,gillingham2002}. A schematic of the experimental set up is shown in figure 1(A), figure 1(B)-(D) outlines the mechanism by which the nanowires are created. When macroscopic wires make contact with each other the atoms on the surface bond to atoms in the other wire (figure 1(B)).  As the wires begin to separate, filaments are formed between the two wires (figure 1(C)). As the wires are separated further, the filaments are stretched and get thinner - in the nanoscale regime, as the wire gets thinner conduction channels close and the conductance falls in steps.  Towards the end of this process, there are only a very small number of atoms remaining in contact between the wires. At the end of the process the wires break (figure 1(D)). The reverse process can occur giving rise to quantized conductance while the macroscopic wires are coming together.

The Cu wires have a 250 $\mu$m diameter, and purity of better than 99.99 \%, and were vibrated at a frequency in the range of 0.5 to 1 Hz.   A bias voltage of 20 mV was connected across the two wires and the current flowing through the wires (about 1 $\mu$A per open conduction channel) was amplified by a transimpedance amplifier. Both the bias and the current were captured by a Tektronix TDS430A digital oscilloscope.  Nanowires were not created every time the wires came together so the data sets had to be filtered to separate those which demonstrated quantum conduction.  A magnetic field was applied perpendicular to the nanowires via a pair of Helmholtz coils that were able to produce $\pm$5 mT - as shown by the arrows in figure 1(A).  

\begin{figure}
\begin{center}
    \includegraphics[width=5.3cm]{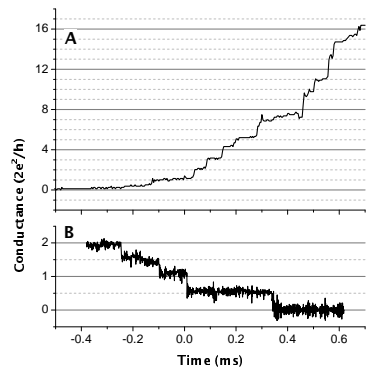}
    \caption{These are representative conductance vs.\ time curves, taken from Cu at room temperature in air at zero applied field.  The different signs of the curves gradient represents that (A) was taken from when the macroscopic Cu wires were coming together and (B) when they were separating.  (A) shows the expected spin degenerate behaviour i.e. the conductance quantized with units of $2e^2/h$, (B) however shows a different behaviour the conductance is quantized with units of $e^2/h$.}
\end{center}
\end{figure}
Figure 2 shows two typical conductance vs.\ time curves obtained with no applied magnetic field, these illustrate 2 points. Firstly that quantum conductance can be seen in nanowires formed by the macroscopic wires: coming together, compressing the nanowires (figure 2(A)) and separating, stretching the nanowires (figure 2(B)). In general the quantization modes are the same for the nanowires formed by compression and tension, apart from the gradient of the curves i.e. when the macroscopic wires are coming together the conductance starts low and ends high and vice versa when the macroscopic wires are separating.  The second point considers the size of the steps in the quantized conduction.  Cu is a diamagnetic material, so from the Landauer theory presented earlier we expect the conductance quantum to be $G_0=2e^2/h$, that is that the spin channels are degenerate.  In figure 2(A) we see that the conductance quantum is indeed $2e^2/h$.  However when we inspect figure 2(B) we see a clear $e^2/h$ quantization.  These two conduction modes are seen both when the macroscopic wires are coming together or separating - so whether the nanowire is in compression or tension has no major effect on the mechanism causing the $e^2/h$ quantization mode.  From inspecting individual conductance vs.\ time curves it is apparent that the total collision time between the macroscopic wires was typically 5--50 ms.

\begin{figure}\begin{center}
    \includegraphics[width=5.3cm]{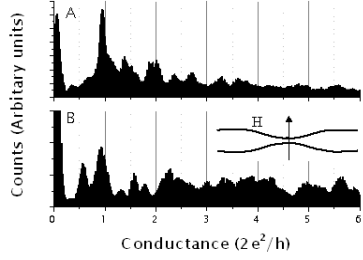}
    \caption{These are representative conduction histograms.  It is built up from conductance vs. time curves taken at room temperature in air. (A) was taken at zero applied magnetic field - shows that the expected $2e^2/h$ quantization is dominant - however there is a small peak at $0.5 \times 2e^2/h$ and $1.5 \times 2e^2/h$ showing that there is some quantization in units of $e^2/h$ occurring. (B) was taken with a 0.5 mT magnetic field applied perpendicular to the nanowire.  This shows that the $e^2/h$ quantization has become far more dominant than in (A) the zero field case.  The inset is a schematic of the field direction with respect to the nanowire.}
\end{center}\end{figure}

Figure 3 shows two representative conductance histograms.  These are built up by counting the number of points in conductance vs.\ time curves which have a particular value of the conductance, thus stable conductance plateaux will show up as peaks in these histograms.  So the peaks in the conduction histograms correspond to the values of quantized conductance. Since each conductance vs.\ time curve represents the creation of a nanowire, these histograms represent an average over many nanowires. Figure 3(A) was taken in zero applied field: as can be readily seen, peaks appear at odd multiples of $e^2/h$ but they are small compared to the peaks at even multiples of $e^2/h$. Figure 3(B) however was taken at a field of about 0.5 mT applied perpendicular to the nanowire (and parallel to one of the macroscopic wires) as shown in the inset of figure 3B. The $e^2/h$ peak is now comparable in strength with the $2e^2/h$ peak.  

\begin{figure}\begin{center}
    \includegraphics[width=10cm]{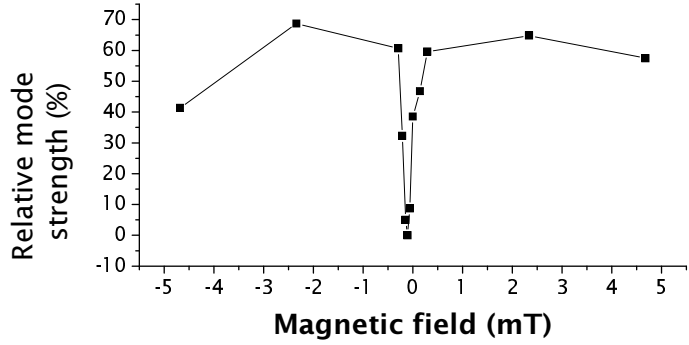}
    \caption{The magnetic field dependence of the relative strength of the $e^2/h$ mode derived from the height of the $e^2/h$ peak from conductance histograms normalized by the number of curves in the conductance histogram.  The maximum is 70\% at 2mT.  All measurements were made at room temperature.  The field was applied parallel to the earth's magnetic field and was perpendicular to the nanowire and parallel to one of the macroscopic wires - as shown in figures 1 and 3.}
\end{center}\end{figure}

Figure 4 shows the magnetic field dependence of the relative strength of the $G=e^2/h$ mode, which is defined as:
\begin{equation}
\mathrm{MC} = \frac{N(\frac{e^2}{h},H)-N(\frac{e^2}{h},0)}{N(\frac{e^2}{h},0)}
\end{equation}
In figure 4, $N(e^2/h,H)$ corresponds to the height of the $e^2/h$ peak at a magnetic field H obtained from a conductance histogram normalized to the number of curves that made up the histogram.  This value corresponds to the stability of the contacts which give rise to this conductance value.  We find a 70 \% magneto-conductance effect for H = 2 mT.  In the bulk Cu is a diamagnet, so this result is unexpected and also surprisingly large.  For a conductance of $e^2/h$ a single conduction channel will be open, this is widely accepted as corresponding to the smallest possible contacts - generally a single atom (or chain of atoms).  In our work the two quantization modes were both seen, with $G_0 = e^2/h$ and $G_0 = 2e^2/h$, but most importantly the relative strengths of these modes were observed to change with the applied magnetic field.  

We next need to consider possible physical mechanisms which explain our principle findings namely that: the conductance quantum is either $G_0 = e^2/h$ or $G_0 = 2e^2/h$ and  a magnetic field dependence clearly exists.  The possible mechanisms we shall consider are based upon either (i) electron scattering due to disorder or impurities in the nanowire or (ii) the creation of a spin-polarized density of states (DOS) in the nanowire.

Structural disorder can cause elastic scattering of electrons to occur in the nanowire.  Brandbyge et al.\ \cite{Brandbyge1997a} found that corrugated edges and localized scatters moved the peaks by an arbitrary amount - and so cannot cause the fixed $e^2/h$ mode.  de Heer et al.\ \cite{deheer1997} found that elastic scatterers can split the nanowire into two nanowires in series - so the conductance will be reduced.  However to obtain the $e^2/h$ mode the two nanowires created by this method would need to have the same number of open conduction channels and it is not clear why this should arise. These mechanisms also do not explain the magnetic field dependence, although it has been shown that the magnetic field can increase the disorder in Au nanowires \cite{manders2001}: in that case a magnetic field strength of 10 T was required which is much larger than that used in this work.

Spontaneous magnetization of nanowires has been predicted in a range of simple non-magnetic metals (Na, Cs and Al \cite{zabala2002}).  At critical radii these 3 dimensional nanowires are predicted to undergo a transition to a spin polarized magnetic state.  Cu like Na and Cs is characterized by s electrons at the Fermi energy, so a spontaneous magnetization could arise within the Cu nanowire over a sufficiently large region to be thermally stable.  Such a spontaneous magnetization will polarize the conduction electrons, and when an external magnetic field, sufficient to align the spin-polarized region in the direction of the field is applied, it will lift the spin degeneracy between the spin up and spin down conduction channels: consequently the conduction mode will change from $2e^2/h$ to $e^2/h$. Similarly an applied field is required for the observation of the $e^2/h$ quantized conductance mode for nanowires made from bulk ferromagnets\cite{ono1999}.  However to our knowledge no conclusive experimental results on spontaneous magnetization formation in nanowires have been reported to date.

It is well known that oxygen can strongly modify the spin-polarized band structure in a wide range of materials (e.g.\ half metallic behaviour arises in the case of some metallic oxides) and that adsorbates can affect the mechanical and electrical properties of nanowires\cite{bahn2002}.  Since this work was performed in air, the macroscopic Cu wires will have inevitably adsorbed oxygen. The adsorbed oxygen atoms can form part of the nanowires - and may become polarized when a magnetic field is applied.  Oxygen can develop a spin-split p-band at the Fermi energy, which has the effect of blocking one of the spin conduction channels and gives rise to the observed magneto-conductance effect. Absorbed oxygen is also one of the possible explanations for the large magneto-conductance effect seen in electrodeposited Ni nanowires \cite{papanikolaou2002a}.  In our work the magneto-conductance effect has been studied in the region when the nanowire will be at its thinnest and so single adsorbed impurities such as oxygen can easily play a role and give rise to the effect we observe.  The macroscopic Cu wires used were of high purity (the highest content of ferromagnetic impurity is that of Fe at 1 ppm).  This means that unless some surface segregation effect concentrates the magnetic impurities at the surface it is unlikely that magnetic atoms will take part in forming sufficient nanowires to cause the observed effect, whether by spin filtering \cite{Papanikolaou2002} or any other mechanism.  In our view the oxygen adsorbate mechanism is the most likely.

To summarize we have seen a significant magneto-conductance effect in Cu nanowires in fields of 2 mT.  We interpret this as being due to the spin filtering effect introduced by oxygen adsorbates modifying the electronic band structure in the nanowire.  In this work surprisingly large spin polarized quantum conduction effects have been seen in a material, which in the bulk is non-magnetic, by reducing the dimensions of the active material: this may offer a new approach in developing spintronic devices, differing from the current research effort devoted to developing bulk materials which have 100 \% spin polarization. We conjecture that this effect could be seen in other non-magnetic metals  when formed into nanowires under the correct conditions.

\ack
The authors would like to thank  Dr WF Egelhoff Jr, Dr EY Tsymbal and Professor A Howie for valuable discussions and the EPSRC for funding the project.

\end{document}